\documentclass[aps,twocolumn,floatfix,groupedaddress,superscriptaddress,longbibliography,footinbib,notitlepage,showpacs]{revtex4-1}
\usepackage{graphicx,graphics,times,bm,bbm,bbold,amssymb}
\usepackage[T1]{fontenc}
\usepackage{amsthm,amsmath,amsfonts,dsfont,color,xcolor,dcolumn}
\usepackage[bookmarksnumbered,colorlinks,plainpages]
{hyperref}
\hypersetup{
     colorlinks = true,
     citecolor  = blue,
     urlcolor   = blue}
\newtheorem{manualtheoreminner}{Theorem}
\newenvironment{manualtheorem}[1]{%
  \renewcommand\themanualtheoreminner{#1}%
  \manualtheoreminner
}{\endmanualtheoreminner}
\newtheorem{manualremarkinner}{Remark}
\newenvironment{manualremark}[1]{%
  \renewcommand\themanualremarkinner{#1}%
  \manualremarkinner
}{\endmanualremarkinner}

\newtheorem{manualpropositioninner}{Proposition}
\newenvironment{manualproposition}[1]{%
  \renewcommand\themanualpropositioninner{#1}%
  \manualpropositioninner
}{\endmanualpropositioninner}
\newtheorem{manualcorollaryinner}{Corollary}
\newenvironment{manualcorollary}[1]{%
  \renewcommand\themanualcorollaryinner{#1}%
  \manualcorollaryinner
}{\endmanualcorollaryinner}

\usepackage{soul}
\begin{document}

\title{Fine-Structure Classification of Multiqubit Entanglement by Algebraic Geometry}

\author{Masoud Gharahi}
{\email[]{masoud.gharahi@gmail.com}
\affiliation{School of Science and Technology, University of Camerino, 62032 Camerino, Italy}
\affiliation{INFN Sezione di Perugia, 06123 Perugia, Italy}
\author{Stefano Mancini}
\affiliation{School of Science and Technology, University of Camerino, 62032 Camerino, Italy}
\affiliation{INFN Sezione di Perugia, 06123 Perugia, Italy}

\author{Giorgio Ottaviani}
\affiliation{Department of Mathematics and Computer Science ``Ulisse Dini'',  University of Florence, 50134 Florence, Italy}

\begin{abstract}
We present a fine-structure entanglement classification under stochastic local operation and classical \mbox{communication} (SLOCC) for multiqubit pure states. To this end, we employ specific algebraic-geometry tools that are SLOCC invariants, secant varieties, to show that for $n$-qubit systems there are $\lceil\frac{2^{n}}{n+1}\rceil$ entanglement families. By using another invariant, $\ell$-multilinear ranks, each family can be further split into a finite number of subfamilies. Not only does this method facilitate the classification of multipartite entanglement, but it also turns out to be operationally meaningful as it quantifies entanglement as a resource.
\end{abstract}


\maketitle

\section{Introduction}
Classification, intended as the process in which ideas and objects are recognized, differentiated, and understood, plays a central role in natural sciences \cite{Wilkins-Ebach}. Adhering to \mbox{mathematics}, classification is collecting sets which can be \mbox{unambiguously} defined by properties that all its members share. As such it becomes a fundamental milestone for \mbox{characterizing} \mbox{\emph{entanglement}} \cite{HHHH09}. As entangled states are a \mbox{basis} for \mbox{quantum-enhanced} applications (see, e.g., Ref. \cite{WGE17}), it becomes of key importance to know which of these states are equivalent in the sense that they are capable of performing the same tasks almost equally well. Finding such equivalence classes, that will provide an entanglement classification based on a finite number of entanglement families, is a long-standing open problem in quantum information theory \cite{HHHH09}.

Having quantum correlations shared by spatially \mbox{separated} parties, the most general local operations that can be implemented, without deteriorating them, are describable by stochastic local operations and classical \mbox{communication} (SLOCC). Thus, it seems natural to seek a finite \mbox{entanglement} classification under SLOCC. Two multiqubit states are SLOCC equivalent if one can be obtained with nonzero \mbox{probability} from the other one using local invertible \mbox{operations}. On the grounds of group theory, SLOCC \mbox{equivalence} classes are orbits under the action of special linear group ${\rm{SL}}(2,\mathbbm{C})^{\times{n}}$ on the set of $n$-qubit states.

SLOCC classification works well for two and three qubits which feature two and six orbits, respectively. However, \mbox{already} for four (or more) qubits, there are infinitely many (\mbox{actually} uncountable) SLOCC classes \cite{DVC00}. This issue has been solved for four qubits, the case which attracted most \mbox{attention} \cite{VDDV02, CD07, CW07, BDDMR10, BK12, CDGZ13, GA16}, and also for $n$-qubit symmetric states \cite{BKMGLS09, RM11}. Although the general case of $n$-qubit entanglement has been addressed, its classification suffers from family \mbox{overlapping} \cite{LL12,GM18}, or still shows an infinite number of classes \cite{GW13}. Thus, it necessitates new methods to establish a finite classification. 

Formally, (pure) quantum states are rays in a Hilbert space. As a consequence, the space of states is more \mbox{appropriately} described by projective Hilbert space $\mathbbm{P}(\mathcal{H}_{n})$. Thus, a \mbox{natural} way to study entanglement of pure states is with algebraic \mbox{geometry}, which is the ``language'' of projective spaces. This avenue was put forward in Refs. \cite{Miyake03, BH01, ST13}, where the \mbox{authors} investigated the geometry of entanglement and considered small systems (up to $\mathbbm{C}^3\otimes\mathbbm{C}^2\otimes\mathbbm{C}^2$) to lighten it. \mbox{Following} this, it has been recently realized the existence, for four qubit systems, of \emph{families}, each including an infinite \mbox{number} of SLOCC classes with common properties \cite{HLT14-17, SBSE17, SMKKKO18}. The \mbox{framework} of algebraic geometry also helped to visualize \mbox{entanglement} families with polytopes \cite{WDGC13, SOK14}, which would be of practical use if a finite classification existed.
 
In this paper, we introduce an entanglement classification of ``generic'' $n$-qubit pure states under SLOCC that is based on a finite number of \emph{families} and \emph{subfamilies} (i.e., a fine-structure classification). We do this by employing tools of algebraic geometry that are SLOCC invariants. In particular, the families and subfamilies will be identified using $k$-secants and $\ell$-multilinear ranks (hereafter $\ell$-multiranks), respectively. A $k$-secant of a variety $\mathcal{X}\subset\mathbbm{P}(\mathcal{H}_{n})$ is the projective span of $k$ points of $\mathcal{X}$. Geometrically, the $k$-secant variety is the zero locus of a set of polynomial equations. Physically, as the \mbox{$k$-secant} of a variety joins its $k$ points, it can liaise to the concept of quantum superposition. On the other hand, \mbox{$\ell$-multiranks} are a collection of integers which are just ranks of different matricizations of a given $n$-qubit state as an order-$n$ tensor in $\mathbbm{C}^{2^{\otimes{n}}}$. Actually, the $\ell$-multiranks tell us about the separability of such a state; when all of them are equal to one we are dealing with a fully separable state. Furthermore, each $k$-secant is a counterpart of the generalized Schmidt rank \cite{CDS08, CCDJW10} which is an entanglement measure. These connections make our classification also operationally meaningful.

\section{The main result}
Algebraic geometry studies projective varieties, which are the subsets of projective spaces defined by the vanishing of a set of homogeneous polynomials, endowed with the structure of algebraic variety. This moved on from studying properties of points of plane curves resulting as solutions of set of polynomial equations (which include lines, circles, parabolas, ellipses, hyperbolas, cubic curves, etc.). Actually, much of the development of algebraic geometry occurred by emphasizing properties that not depend on any particular way of embedding the variety in an ambient coordinate space. This was obtained by extending the notion of point. In this framework, the Segre embedding is used to consider the Cartesian product of projective spaces as a projective variety. This takes place through the map
\begin{equation}\nonumber
\Sigma_{(d_1-1,d_2-1)}^{2}:~\mathbbm{P}^{d_1-1}\times\mathbbm{P}^{d_2-1}\hookrightarrow\mathbbm{P}^{d_1d_2-1}\, ,
\end{equation}
which takes a pair of points $([x],[y])\in\mathbbm{P}^{d_1-1}\times\mathbbm{P}^{d_2-1}$ to their products $([x_{0}:x_{1}:\cdots:x_{d_1-1}],[y_{0}:y_{1}:\cdots:y_{d_2-1}]) \rightarrow [x_{0}y_{0}:x_{0}y_{1}:\cdots:x_{i}y_{j}:\cdots:x_{d_1-1}y_{d_2-1}]\,$, where the notation refers to homogeneous coordinates and the $x_{i}y_{j}$ are taken in lexicographical order. The image of this map is called Segre variety.

Now, let us consider an $n$-qubit state:
\begin{equation}\label{qubitstate}
|\psi\rangle = \sum_{i\in\{0,1\}^{n}}\mathfrak{c}_{i}|i\rangle\,.
\end{equation}
The space of states $|\psi\rangle$ that are fully separable has the structure of a Segre variety \cite{Miyake03, Heydari08} which is embedded in the ambient space as follows:
\begin{equation}\label{segre}
\Sigma^{n}_{\textbf{1}}:~\mathbbm{P}^{1}\times\mathbbm{P}^{1}\times\cdots\times\mathbbm{P}^{1}\hookrightarrow\mathbbm{P}^{2^{n}-1}\,,
\end{equation}
where $\textbf{1}=(1,\ldots,1)$ and $\times$ is the Cartesian product of sets.
A $k$-secant of the Segre variety joins its $k$ points, each of which represents a distinct separable state. Thus, the joining of points corresponds to an entangled state being a superposition of $k$ separable states. The union of $k$-secant of the Segre variety $\Sigma^{n}_{\textbf{1}}$ gives rise to the $k$-secant variety $\sigma_k(\Sigma^{n}_{\textbf{1}})$. This is as much as the set of entangled states arising from the superposition of $k$ separable states.
Since $k$-secant varieties are SLOCC invariants (see Appendix \ref{AppA}), SLOCC classes congregate naturally into entanglement families. Therefore, the dimension of the higher $k$-secant, which fills the projective Hilbert space of $n$ qubits, can indicate the number of entanglement families.
The higher secant varieties in $\mathbbm{P}(\mathbbm{C}^{2^{\otimes{n}}})$, have the expected dimension
$${{\rm{dim}}}~\sigma_{k}(\Sigma^{n}_{\textbf{1}})= \min\{k(n+1)-1, 2^{n}-1\}\,,$$
for every $k$ and $n$, except $\sigma_{3}(\Sigma^{4}_{\textbf{1}})$ which has dimension $13$ \cite{CGG11}. Consequently, the $k$-secant fills the ambient space, when $k=\lceil\frac{2^{n}}{n+1}\rceil$.
This $k$ indicates the number of entanglement families which remains finite (although growing exponentially) with the number of qubits.

The proper $k$-secant (the states that belongs to $k$-secant but not to ($k-1$)-secant), i.e., the set $\sigma_{k}(\Sigma^{n}_{\textbf{1}})\setminus\sigma_{k-1}(\Sigma^{n}_{\textbf{1}})$, is the union of the $k$-secant hyperplanes $\mathcal{S}_{k}\subset\sigma_{k}(\Sigma^{n}_{\textbf{1}})$ represented by
\begin{equation}\label{G-secant}
\mathcal{S}_{k}=\sum_{i=1}^{k}\lambda_{i}p_{i}\,,
\end{equation}
with $\{\lambda_{i}\}_{i=1}^{k}\neq{0}$ and each $p_{i}$ is a distinct point in $\Sigma^{n}_{\textbf{1}}$.

It is worth saying that each secant, with regards to its dimension, could have tangents as its closure (see  Appendix \ref{AppA}) which discriminate subfamilies with the same $\ell$-multiranks and provide us exceptional states \cite{ST13}. Let us now consider the limits of secants to obtain the tangents. Let $(i_{1},i_{2},\ldots,i_{k})$ be a rearrangement of points indices in Eq. (\ref{G-secant}). The first limit type is when one point tends to another one, i.e., $p_{i_{2}}\to{p_{i_{1}}}$, and let us call the result $p'_{i_{1}}$. The second limit type can be considered as the closure of the first limit type so the third point is  approaching $p_{i_{1}}+\eta p'_{i_{1}}$. The third limit type can be considered as the closure of the second limit type so two points tend to $p_{i_{1}}$ and $p_{i_{2}}$ (if the join of $p_{i_{1}}$ and $p_{i_{2}}$ is still in $\Sigma^{n}_{\textbf{1}}$) \cite{BL14}. As we can always redefine Eq. (\ref{G-secant}) to have the desired form and new coefficients rather than $\lambda_{j}$, we can formulate these limits as
\begin{align}\label{G-tangent-g1}
T^{(1)}_{k}=&\lim_{\epsilon\to{0}}\frac{\lambda_{i_{2}}}{\epsilon}\big(p_{i_{2}}(\epsilon)-p_{i_{1}}\big)+\sum_{j=i_{3}}^{i_{k}}\lambda_{j}p_{j}\,,
\\ \label{G-tangent-g2}\nonumber
T^{(2)}_{k}=&\mu_{1}p'_{i_{1}}+\lim_{\eta\to{0}}\frac{\mu_{2}}{\eta^{2}}\big(p_{i_{3}}(\eta)-(p_{i_{1}}+\eta\, p'_{i_{1}})\big) \\
&+\sum_{j=i_{4}}^{i_{k}}\lambda_{j}p_{j}\,,~~~
\\ \label{G-tangent-g3}\nonumber
T^{(3)}_{k}=&\lim_{\epsilon\to{0}}\frac{\nu_{1}}{\epsilon}\big(p_{i_{3}}(\epsilon)-p_{i_{1}}\big)+\lim_{\epsilon\to{0}}\frac{\nu_{2}}{\epsilon}\big(p_{i_{4}}(\epsilon)-p_{i_{2}}\big) \\
&+\sum_{j=i_{4}}^{i_{k}}\lambda_{j}p_{j}\,.
\end{align}
Obviously, these processes can be generalized if we consider all extra limit types which may occur by adding the next points. This will provide us higher tangential varieties.

On the other hand, $\ell$-multiranks are ${n\choose\ell}$-tuples of ranks of matrices which can be obtained by tensor flattening (or matricization) \cite{Landsberg}. Not only do the integers of the tuples tell us about the separability of the state (each integer equals one means there is a separability between two parties) but also the greater the integers are, the more entanglement the parties of the state have. In addition, as $\ell$-multiranks are also SLOCC invariants (see Appendix \ref{AppA}), the SLOCC classes in each family gather into subfamilies.

Therefore, we use $k$-secant varieties and $\ell$-multiranks as the SLOCC invariants to group orbits (classes) into finite number of families and subfamilies. In addition, one can split $k$-secant families, according to Theorem \ref{theorem1} in Appendix \ref{AppA}, by identifying their closure as $k$-tangent. Hence, the classification algorithm can be summarized as:
{\bf (i)} find families by identifying $\Sigma^{n}_{\textbf{1}}$, $\sigma_{2}(\Sigma^{n}_{\textbf{1}}), \ldots, \sigma_{k}(\Sigma^{n}_{\textbf{1}})$, 
{\bf (ii)} split families to secants and tangents by identifying $\tau_{2}(\Sigma^{n}_{\textbf{1}}), \ldots, \tau_{k}(\Sigma^{n}_{\textbf{1}})$, and 
{\bf (iii)} find subfamilies by identifying $\ell$-multiranks.

\section{Examples}
($n=2$). Classification of two-qubit states is fairly trivial, nonetheless it can be instructive for working out the developed concepts. For the Segre surface $\Sigma^{2}_{\textbf{1}}$, we shall use homogeneous coordinates associated with the induced basis $\left\{|00\rangle,|01\rangle,|10\rangle,|11\rangle\right\}$. That is to say, a point $p\in\mathbbm{P}^3$ is written in homogenous coordinates $\left[\mathfrak{c}_0:\mathfrak{c}_1:\mathfrak{c}_2:\mathfrak{c}_3\right]$ whenever $p$ is the projective class of the two-qubit state of Eq. (\ref{qubitstate}). Then, the Segre surface $\Sigma^{2}_{\textbf{1}}$ is the projective variety with points given by affine coordinates $[1:a:b:ab]$, where $a$ and $b$ are complex parameters. This expression must be properly understood, in that the limits of $a$ and/or $b$ going to  infinity, must be included. It is easy to see that $|\Phi^{\pm}\rangle=[1:0:0:\pm 1]$ and $|\Psi^{\pm}\rangle=[0:1:\pm 1:0]$ (the well-known Bell states) are elements of $\sigma_{2}(\Sigma^{2}_{\textbf{1}})$ which is given by Eq. (\ref{G-secant}). Considering $p_{2}(\epsilon)=[1:a_{1}+\epsilon:b_{1}+\epsilon:(a_{1}+\epsilon)(b_{1}+\epsilon)]$ and using Eq. (\ref{G-tangent-g1}) to create the closure of the two-secant, we have the special situation that all points on the tangent lines $T^{(1)}_{2}$ lie also on two-secant. It means that all elements of $\mathbbm{P}^{3}$ are elements of $\sigma_{2}(\Sigma^{2}_{\textbf{1}})$. One can thus conclude that all entangled states of two qubits are linear combinations of two separable states, which is the same result obtainable by the Schmidt decomposition. Here the two entanglement families coincide with the two SLOCC classes, namely, separable and entangled. 

Already from this example we can draw a general conclusion. That is, for $n\geq 2$ we have
\begin{equation}\label{general2}
\mathcal{P}\{|{\rm{Bell}}\rangle|\text{1-qubit}
\rangle^{\otimes{(n-2)}}\}\in\sigma_{2}(\Sigma^{n}_{\textbf{1}})\,,
\end{equation}
where $\mathcal{P}\{\cdot\}$ denotes all possible permutations.

($n=3$). For three qubits the Segre three-fold $\Sigma^{3}_{\textbf{1}}\subset\mathbbm{P}^{7}$ consists of general points $[1:a:b:ab:c:ac:bc:abc]$ with the possibility of $a$ and/or $b$ and/or $c$ going to infinity. Moving on to the proper two-secant variety, we have generic elements as $[\lambda_{1}+\lambda_{2}:\lambda_{1} a_{1}+\lambda_{2} a_{2}:\lambda_{1} b_{1}+\lambda_{2} b_{2}:\lambda_{1} a_{1}b_{1}+\lambda_{2} a_{2}b_{2}: \lambda_{1} c_{1}+\lambda_{2} c_{2}:\lambda_{1} a_{1}c_{1}+\lambda_{2} a_{2}c_{2}:\lambda_{1} b_{1}c_{1}+\lambda_{2} b_{2}c_{2}:\lambda_{1} a_{1}b_{1}c_{1}+\lambda_{2} a_{2}b_{2}c_{2}]$. One can check that $|{\rm{GHZ}}_{3}\rangle=[1:0:0:0:0:0:0:1]$ is an element of $\sigma_{2}(\Sigma^{3}_{\textbf{1}})$. We also need to consider situations in which one or more parameters tend to infinity. As an example, let us take $a_{1}=b_{1}=\sqrt{\lambda_{2}}\to\infty$ with $c_{1}=c_{2}$, which gives the biseparable state $|{\rm{B}}_{A-BC}\rangle=[1:a:b:c:d:ad:bd:cd]$. Hence, the state $|{\rm{GHZ}}_{3}\rangle$ with one-multirank equal to $(222)$ and all three biseparable states $|{\rm{B}}_{i}\rangle_{i=1}^{3}$ with the same form as Eq. (\ref{general2}) and one-multiranks equal to $(122)$, $(212)$, and $(221)$, are elements of $\sigma_{2}(\Sigma^{3}_{\textbf{1}})$. However, the tangent points defined in Eq. \eqref{G-tangent-g1} cannot be expressed as elements of $\sigma_{2}(\Sigma^{3}_{\textbf{1}})$, which spans all $\mathbbm{P}^{7}$ only if the tangential variety is included as its closure. If we consider the tangent to $p_{1}=[1:0:0:0:0:0:0:0]$ (equivalent to all points on $\Sigma^{3}_{\textbf{1}}$ by a SLOCC), we have $T^{(1)}_{2}=[1:\lambda:\lambda:0:\lambda:0:0:0]\in\tau_{2}(\Sigma^{3}_{\textbf{1}})$ [e.g., $|{\rm{W}}_{3}\rangle=\lim_{\lambda\to\infty}T^{(1)}_{2}$ with one-multirank equal to $(222)$]. We saw that one-multirank equal to $(222)$ can be discriminated by secant and/or tangent classification. From now on, we use a prime for the states in tangent to discriminate secant and tangent families where they have same $\ell$-multiranks. In summary, this classification provides us two secant families (three secant/tangent families), and six subfamilies (Table \ref{table:1}, see also Ref. \cite[Example 14.4.5]{GKZ}) that coincide with the six SLOCC classes of Ref. \cite{DVC00}.

Also from this example we can extrapolate general results. That is, for $n\geq{r}\geq{3}$, we have
\begin{align}\label{general3}\nonumber
\hspace{8mm}&|{\rm{GHZ}}_{n}\rangle=|0\rangle^{\otimes{n}}+|1\rangle^{\otimes{n}} &\in \sigma_{2}(\Sigma^{n}_{\textbf{1}})\,,\hspace{8mm} \\ \nonumber
&\mathcal{P}\{|{\rm{GHZ}}_{r}\rangle|\text{1-qubit}\rangle^{\otimes{(n-r)}}\} &\in \sigma_{2}(\Sigma^{n}_{\textbf{1}})\,,\hspace{8mm} \\ \nonumber
&|{\rm{W}}_{n}\rangle=|{\rm{D}}_{n}^{1}\rangle &\in \tau_{2}(\Sigma^{n}_{\textbf{1}})\,,\hspace{8mm} \\
&\mathcal{P}\{|{\rm{W}}_{r}\rangle|\text{1-qubit}\rangle^{\otimes{(n-r)}}\} &\in \tau_{2}(\Sigma^{n}_{\textbf{1}})\,,\hspace{8mm}
\end{align}
where
$$|{\rm{D}}_{n}^{l}\rangle={\binom{n}{l}}^{-(1/2)}\sum_{i}\mathcal{P}_{i}\{|0\rangle^{\otimes{(n-l)}}\otimes|1\rangle^{\otimes{l}}\}\,,$$
are the so-called Dicke states (with $l$ excitations)

($n=4$). Due to Remark \ref{remark2} and Corollary \ref{corollary} in Appendix \ref{AppA} and classification of two- and three-qubit states, we have (1) all triseparable states $|{\rm{T}}_{i}\rangle_{i=1}^{6}$ from Eq. (\ref{general2}) are elements of $\sigma_{2}(\Sigma^{4}_{\textbf{1}})$, (2) all biseparable states $|{\rm{B}}_{i}^{{\rm{GHZ}}_{3}}\rangle_{i=1}^{4}$ and $|{\rm{B}}_{i}^{{\rm{W}}_{3}}\rangle_{i=1}^{4}$ from Eq. (\ref{general3}) are, respectively, elements of $\sigma_{2}(\Sigma^{4}_{\textbf{1}})$ and $\tau_{2}(\Sigma^{4}_{\textbf{1}})$, and (3) the states $|{\rm{GHZ}}_{4}\rangle$ and $|{\rm{W}}_{4}\rangle$ are elements of $\sigma_{2}(\Sigma^{4}_{\textbf{1}})$ and $\tau_{2}(\Sigma^{4}_{\textbf{1}})$, respectively. The rest of the subfamilies of four-qubit states can be identified by considering the elements of three- and four-secants and their closures. The proper three-secant, i.e., the set $\sigma_{3}(\Sigma^{4}_{\textbf{1}})\setminus\sigma_{2}(\Sigma^{4}_{\textbf{1}})$, is the union of the secant hyperplanes $\mathcal{S}_{3}$ represented by Eq. (\ref{G-secant}). For instance, $\alpha|0000\rangle+\beta|0011\rangle+\gamma|1111\rangle$, which comes from joining $|{\rm{GHZ}}_{4}\rangle$ and an element of $\Sigma^{4}_{\textbf{1}}$,
is an element of $\sigma_{3}(\Sigma^{4}_{\textbf{1}})$. To construct the closure of $\sigma_{3}$, we consider different limit types as in Eqs. (\ref{G-tangent-g1})-(\ref{G-tangent-g3}) at $p_{1}=[1:0:\cdots:0]$, equivalent to all points on $\Sigma^{4}_{\textbf{1}}$ by a SLOCC. Then, $|{\rm{W}}_{4}\rangle+|1111\rangle$ and $|{\rm{W}}_{4}\rangle+|0011\rangle$ belong to the first limit type, i.e., Eq. (\ref{G-tangent-g1}) while $|{\rm{D}}_{4}^{2}\rangle$ is an element of the second limit type, i.e., Eq. (\ref{G-tangent-g2}). For the third limit type [Eq. (\ref{G-tangent-g3})], one can take $p_{1}=[0:1:0:\cdots:0]$ as a second point, where $\lambda_{1}p_{1}+\lambda_{2}p_{2}\in\Sigma^{4}_{\textbf{1}}$ and hence $|{\rm{W}}_{4}\rangle+\alpha|0011\rangle+\beta|0101\rangle+\gamma|1001\rangle$ can be considered as a representative example. We denote the union of these points as the tangential variety $\tau_{3}(\Sigma^{4}_{\textbf{1}})$. The proper four-secant, i.e., the set $\sigma_{4}(\Sigma^{4}_{\textbf{1}})\backslash\sigma_{3}(\Sigma^{4}_{\textbf{1}})$, is the union of the secant hyperplanes $\mathcal{S}_{4}$ represented by Eq. (\ref{G-secant}). For instance, $|{\rm{Cl}}_{4}\rangle=\frac{1}{2}(|0000\rangle+|0011\rangle+|1100\rangle-|1111\rangle)$, which is known as cluster state \cite{BR01}, is an element of $\sigma_{4}(\Sigma^{4}_{\textbf{1}})$. As another example, all biseparable states $|{\rm{BB}}_{i}\rangle_{i=1}^{3}=|{\rm{Bell}}\rangle|{\rm{Bell}}\rangle$ which are tensor products of two Bell states are also elements of $\sigma_{4}(\Sigma^{4}_{\textbf{1}})$. Since the highest tensor rank for a four-qubit state is 4 \cite{Brylinski02}, we do not need to construct the four-tangent. To have an exhaustive classification, we have written each subfamily of three- and four-secant families in terms of their two-multiranks in Table \ref{table:2} (more details in Appendix \ref{AppB}). An important observation is that, all elements in $\sigma_{3}(\Sigma^{4}_{\textbf{1}})$ are genuinely entangled. This can be useful for characterizing genuine multilevel entanglement when we look at four qubits as two ququarts \cite{KRBHG18}. Briefly, this classification provide us four secant families (six secant/tangent families), and $35$ subfamilies (Table \ref{table:2}). The petal-like classification of SLOCC orbits is presented in Fig. \ref{fig:1}.

\begin{table}[t]
\centering
\caption{Fine-structure classification of three-qubit entanglement.}
\begin{tabular}{ccccccc}
\hline\hline
& & & & & &\\ [-2ex]
$~$ & $\Sigma^{3}_{\textbf{1}}$ & $\,~\qquad\qquad\qquad$ & $\sigma_{2}$ & $\,~\qquad\qquad\qquad$ & $\tau_{2}$ & $~$ \\ [0.5ex]
\hline
& & & & & & \\ [-2ex]
$~$ & $|{\rm{Sep}}\rangle$ & $\,~\qquad\qquad\qquad$ & $|{\rm{GHZ}}_{3}\rangle$ & $\,~\qquad\qquad\qquad$ & $|{\rm{W}}_{3}\rangle$ & $~$ \\ [0.5ex]
$~$ & & $\,~\qquad\qquad\qquad$ & $|{\rm{B}}_{i}\rangle_{i=1}^{3}$ & $\,~\qquad\qquad\qquad$ & & $~$ \\ [0.5ex]
\hline\hline
\end{tabular}
\label{table:1}
\end{table}
\begin{table}[t]
\centering
\caption{Fine-structure classification of four-qubit entanglement.}
\begin{tabular}{cccccc}
\hline\hline
& & & & & \\ [-2ex]
$\Sigma^{4}_{\textbf{1}}$ & $\sigma_{2}$ & $\tau_{2}$ & $\sigma_{3}$ & $\tau_{3}$ & $\sigma_{4}$\\ [0.5ex]
\hline
& & & & & \\ [-2ex]
$|{\rm{Sep}}\rangle$ & $|{\rm{GHZ}}_{4}\rangle$ & $|{\rm{W}}_{4}\rangle$ & $|(333)\rangle$ & $|(333)'\rangle$ & $|(444)\rangle$\\ [0.5ex]
& $|{\rm{B}}_{i}^{{\rm{GHZ}}_{3}}\rangle_{i=1}^{4}$ & $|{\rm{B}}_{i}^{{\rm{W}}_{3}}\rangle_{i=1}^{4}$ & $|(332)\rangle$ & $|(332)'\rangle$
 & $|(443)\rangle$\\ [0.5ex]
& $|{\rm{T}_{i}}\rangle_{i=1}^{6}$ & & $|(323)\rangle$ & $|(323)'\rangle$
 & $|(434)\rangle$\\ [0.5ex]
& & & $|(233)\rangle$ & $|(233)'\rangle$
 & $|(344)\rangle$\\ [0.5ex]
& & & & & $|(442)\rangle$\\ [0.5ex]
& & & & & $|(424)\rangle$\\ [0.5ex]
& & & & & $|(244)\rangle$\\ [0.5ex]
& & & & & $|{\rm{BB}}_{i}\rangle_{i=1}^{3}$\\ [0.5ex]
\hline\hline
\end{tabular}
\label{table:2}
\end{table}

($n\geq{4}$). We can draw the following conclusions for $n\geq{4}$:
\begin{align}\label{general4}\nonumber
\hspace{1mm}&|{\rm{M}}_{n}^{r}\rangle:=|{\rm{GHZ}}_{n}\rangle+\mathcal{P}\{|0\rangle^{\otimes{r}}|1\rangle^{\otimes{(n-r)}}\} &\in \sigma_{3}(\Sigma^{n}_{\textbf{1}})\,,\hspace{10mm} \\ \nonumber
&|0\rangle_{i}|{\rm{GHZ}}_{n-1}\rangle+|1\rangle_{i}\mathcal{P}\{|1\rangle^{\otimes{s}}|0\rangle^{\otimes{(n-s-1)}}\} &\in \sigma_{3}(\Sigma^{n}_{\textbf{1}})\,,\hspace{10mm} \\ \nonumber
&|{\rm{N}}_{n}^{t}\rangle:=|{\rm{W}}_{n}\rangle+\mathcal{P}\{|1\rangle^{\otimes{t}}|0\rangle^{\otimes{(n-t)}}\} &\in \tau_{3}(\Sigma^{n}_{\textbf{1}})\,,\hspace{10mm} \\ \nonumber
&|0\rangle_{i}|{\rm{W}}_{n-1}\rangle+|1\rangle_{i}\mathcal{P}\{|1\rangle^{\otimes{(t-1)}}|0\rangle^{\otimes{(n-t)}}\} &\in \tau_{3}(\Sigma^{n}_{\textbf{1}})\,,\hspace{10mm} \\ \nonumber
&|{\rm{G}}_{n}^{r}\rangle:=\mathcal{P}\{\alpha|0\rangle^{\otimes{n}}+\beta|0\rangle^{\otimes{r}}|1\rangle^{\otimes{(n-r)}} &\hspace{10mm} \\
&\qquad\qquad\quad+\gamma|1\rangle^{\otimes{r}}|0\rangle^{\otimes{(n-r)}}+\delta|1\rangle^{\otimes{n}}\} &\in \sigma_{4}(\Sigma^{n}_{\textbf{1}})\,,\hspace{10mm}
\end{align}
where $2\leq{r}\leq{n-2}$, $1\leq{s}\leq n-2$, $2\leq{t}\leq n$, and $i=1,\ldots,n$. It is worth noting that the state $|{\rm{G}}_{n}^{r}\rangle$ is a generalization of bipartite state $\alpha|00\rangle+\beta|01\rangle+\gamma|10\rangle+\delta|11\rangle$ and its minor is $2|\alpha\delta-\beta\gamma|$, which coincides with the definition of concurrence \cite{Wootters98}. Therefore, if $\alpha\delta\neq\beta\gamma$, the state $|{\rm{G}}_{n}^{r}\rangle$ is genuinely entangled, otherwise it is biseparable (a tensor product of two $r$- and ($n-r$)-partite entangled states).

\begin{figure}
\center{\includegraphics[width=8.7cm]{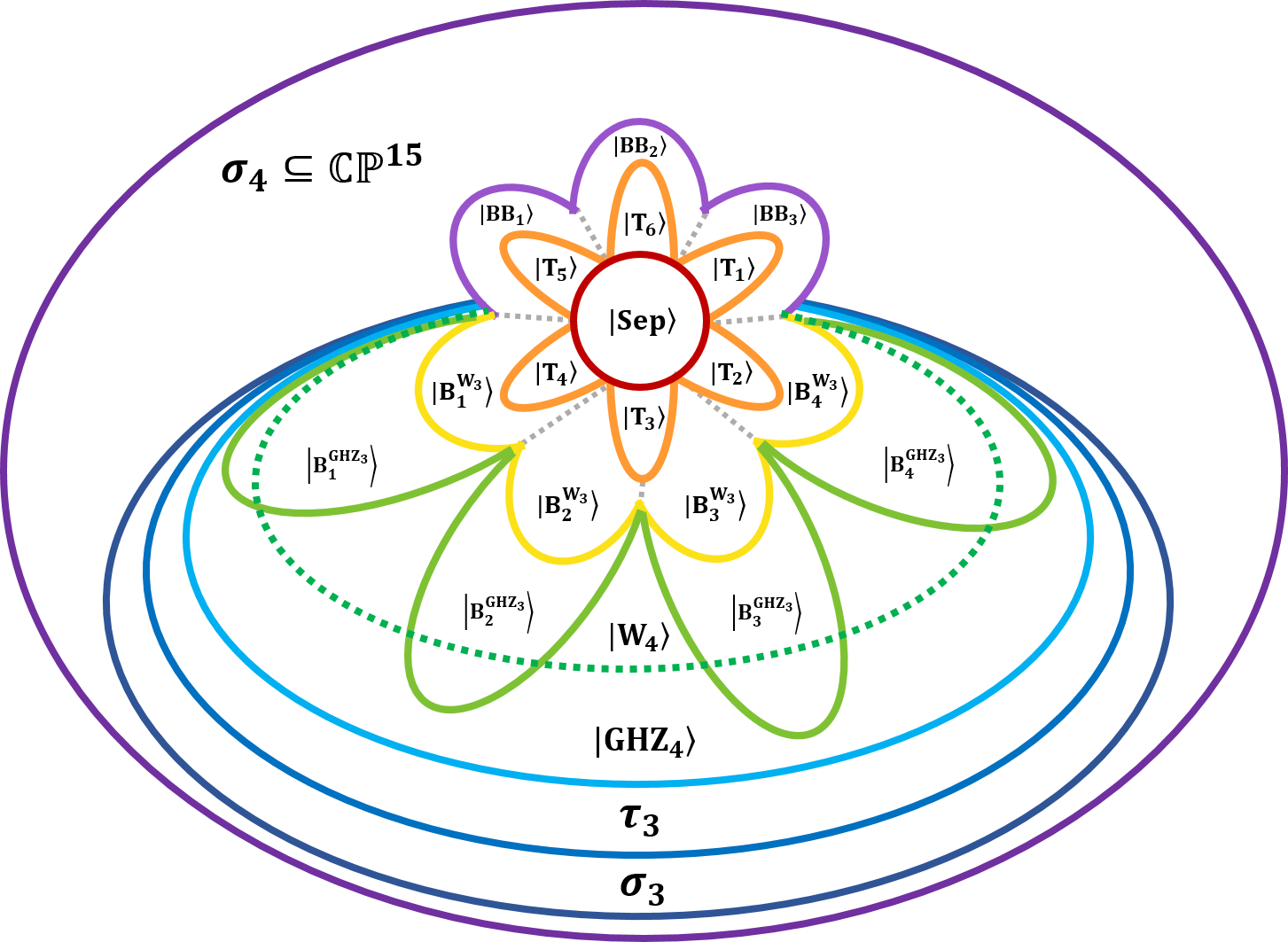}}
\caption{\label{fig:1} (color online). Petal-like classification of SLOCC \mbox{orbits} of four-qubit states. Dashed gray lines in the core show that each $|{\rm{BB}}_{i}\rangle$ encompasses two triseparable subfamilies, while each $|{\rm{B}}_{i}^{{\rm{W}}_{3}}\rangle$ encompasses three triseparable subfamilies. The convex hull of $|{\rm{W}}_{4}\rangle$ (dashed green curve) indicates that this family does not \mbox{encompass} biseparable states $|{\rm{B}}_{i}^{{\rm{GHZ}}_{3}}\rangle$, while both encompass the \mbox{yellow}, \mbox{orange}, and red subsets. From the outer classes, one can go to the inner ones by noninvertible SLOCC (from $\sigma_k$ to $\tau_k$ also in an approximate way), thus generating the entanglement hierarchy. (See Fig. \ref{fig:2} in Appendix \ref{AppB} for more details.)}
\end{figure}

\begin{manualproposition}{1}
For $n\geq{4}$ qubits, there is no symmetric entangled state in the higher secant variety.
\end{manualproposition}

The superposition of $n$-qubit Dicke states with all possible excitations  
\begin{equation}\label{symmetric}
|\psi_{n}^{\rm{Sym}}\rangle=\sum_{l=0}^{n}\mathfrak{d}_{l}|{\rm{D}}_{n}^{l}\rangle\,,
\end{equation}
is the most general symmetric entangled state. The symmetric $n$-qubit separable states have the structure of Veronese variety ($\mathcal{V}^{n}_{1}$) and its $k$-secant varieties are SLOCC families \cite{BH01, ST13, SBSE17}. The higher $k$-secant variety fills the ambient space for $k=\lceil\frac{n+1}{2}\rceil$. Comparing with the higher $k$-secant in Segre embedding ($k=\lceil\frac{2^{n}}{n+1}\rceil$), it proves the proposition. Moreover, we will show below that each Dicke state with $1\leq{l}\leq\lfloor\frac{n}{2}\rfloor$ (the same for the spin-flipped version, i.e., $|{\rm{D}}_{n}^{n-l}\rangle$) is in a $k$-secant family of Veronese embedding, and hence, Segre embedding for $2\leq{k}\leq\lfloor\frac{n}{2}\rfloor+1$, respectively. Thus, this method can be useful to classify entanglement of symmetric states and the corresponding number of families grows slower than Ref. \cite{BKMGLS09}.

Consider the following $n$-qubit separable state:
$$|{\rm{S}}_n(\varepsilon)\rangle=(|0\rangle+\varepsilon|1\rangle)^{\otimes{n}}=\sum_{l=0}^{n}\varepsilon^{l}|{\rm{D}}_{n}^{l}\rangle\,.$$
Thanks to the definition of tangent star and Eqs. (\ref{recursivesecant}) and (\ref{G-tangent}) in Appendix \ref{AppA}, we can write
\begin{equation}\label{tangent-dicke}
\lim_{\varepsilon\to{0}}\frac{1}{\varepsilon^{m+1}}\left(|{\rm{S}}_n(\varepsilon)\rangle-\sum_{i=0}^{m}\varepsilon^{i}|{\rm{D}}_{n}^{i}\rangle\right)=|{\rm{D}}_{n}^{m+1}\rangle\in\tau_{m+2}(\Sigma^{n}_{\textbf{1}})\,,
\end{equation}
where $0\leq{m}\leq\lfloor\frac{n}{2}\rfloor-1$. Furthermore, $\lfloor\frac{n}{2}\rfloor$-multiranks of the Dicke states with $1\leq l\leq\lfloor\frac{n}{2}\rfloor$ (and similarly $|{\rm{D}}_{n}^{n-l}\rangle$) are $l+1=k$ ($\ell$-multiranks with $\ell<\lfloor\frac{n}{2}\rfloor$ have the same value or maximum rank). We guess that this is a general behavior which holds true for symmetric multiqudit systems as well. In a similar way, one can check that the states $|{\rm{N}}_{n}^{r}\rangle$ are on the limiting lines of the states $|{\rm{M}}_{n}^{r}\rangle$ in Eq. (\ref{general4}), and therefore, are exceptional states.

Consider now $|\psi_{4}^{\rm{Sym}}\rangle$ from Eq. (\ref{symmetric}) which belongs to $\tau_{3}(\Sigma^{4}_{\textbf{1}})$. It can asymptotically produce lower tangent elements, like $|{\rm{W}}_{4}\rangle$. The state $|{\rm{W}}_{4}\rangle$ also can be asymptotically produced from the state $|{\rm{M}}_{4}\rangle$ which belongs to $\sigma_{3}(\Sigma^{4}_{\textbf{1}})$ (see Appendix \ref{AppB}). 

\begin{manualremark}{1}\label{remark1}
States living in the higher secant and/or tangent can produce all states in the lower secants and/or tangents by means of degenerations, that is performing some limits.
\end{manualremark}

\section{Conclusion}
We presented a fine-structure entanglement classification that can be interpreted as a Mendeleev table, where the structure of an element can be used as a core structure of another. As a matter of fact, for $n$-qubit classification we are fixing the elements in $k$-secant families [see Eqs. (\ref{general2})-(\ref{general4})], and, indeed, one can always use $n$-qubit classification as a partial classification of ($n+1$)-qubit case. Then, we just need to find the elements of new $k$-secants for the classification of ($n+1$)-qubit states. As we have already illustrated in our examples, new $k$-secants' elements can be identified by joining points of previous $k$-secant families, and considering all tangential varieties (see also Appendix \ref{AppA}). More interesting is that joining randomly chosen elements from both $\sigma_{i}$ and $\sigma_{j}$ would land in $\sigma_{i+j}\setminus\sigma_{i+j-1}$, with probability one \cite{Landsberg}. Therefore, one can always create a general element in a desired secant family. In addition, all the genuine entangled states in higher secants and tangents can be, respectively, considered as the generalizations of ${\rm{GHZ}}$ and ${\rm{W}}$ states in two-secant and two-tangent [one can also see a footprint of ${\rm{GHZ}}$ and ${\rm{W}}$ states in the higher secants and tangents from Eq. (\ref{general4})].

To clearly show the potentialities of our approach, we have elaborated the classification for $n=5$ qubits in Appendix \ref{AppC}. We believe the method can be extended to find a classification of multipartite entanglement for higher dimensional systems as we have already provided a conjecture for the classification of symmetric multiqudit states.

We emphasize the operational meaning of the proposed classification as it somehow measures the amount of entanglement in multipartite systems, where a well-established entanglement monotone is still lacking.
Furthermore, the tools we proposed for entanglement characterization can also be useful as states complexity measures, since they share analogies with the tree size method presented in Refs. \cite{LCWRS14, CLS15}. Indeed, the notion of tree size can be understood as the length of the shortest bracket representation of a state, which in turn is the tensor rank. Additionally, they offer a perspective for evaluating the computational complexity of quantum algorithms, by analyzing how the classes change while running them (see also Ref. \cite{HJN16-JH18}).

Still along the applicative side, since in a system with a growing number of particles, most of the states cannot be realistically prepared and will thus never occur neither in natural nor in engineered quantum systems \cite{WGE17}, our coarse-grain classification could provide a tool to singling out states that we do effectively need (e.g., a representative of each family and/or subfamily). For instance, ${\rm{W}}$ states that are living in a lower secant, although useful for many processes like the realization of quantum memories \cite{LST04}, are known to be more robust but not very entangled. Hence, for other tasks, like quantum teleportation, the usage of ${\rm{GHZ}}$ states that are more entangled has been suggested \cite{ZCZYBP09}, i.e., move up from the tangent to the proper secant of the lower secant family. Indeed ${\rm{GHZ}}$ states provide some degree of precision in frequency measurements \cite{BIWH96}, but in Ref. \cite{HMPEPC97} this is increased (even in the presence of decoherence), using a state lying in higher secant. Hence, it seems that higher secant families offer better estimation accuracy in quantum metrology (see also Refs. \cite{HLKSWWPS12, Toth12}). Also, our results about the cluster state $|{\rm{Cl}}_{4}\rangle$, supports the idea that states living in higher secants are more suitable as a resource for measurement-based quantum computation \cite{BBDRV09}. Actually, going to higher secants makes states more entangled and at the same time also more robust (at least with respect to losses) because even losing one qubit there would always be some residual entanglement left.

Finally, based on our classification, one can construct new entanglement witnesses to be used for detecting entanglement in multipartite mixed states (where state tomography is not efficient). Already, in Ref. \cite{BSV12} it has been shown that one can find, following a geometric approach, device-independent entanglement witnesses that allow us to discriminate between various types of multiqubit entanglement. We believe that this could also pave the way to extend this classification to mixed states, and to study the entanglement depth \cite{LRBPBG15, L-etal-prx-18} of each class.

\section*{Acknowledgments}
M.~G. thanks to the University of the Basque Country for the kind hospitality during the early stage of this work. There, he is grateful to I.~L.~Egusquiza and M.~Sanz for discussing and sharing notes on the subject of the present paper. He also acknowledges delightful and fruitful discussions subsequently had with Jaros\l{}aw Buczy\'{n}ski, Joachim Jelisiejew, \mbox{Pedram} Karimi, and Reza Taleb. G.~O. is a member of \mbox{GNSAGA}.

\appendix
\renewcommand*{\appendixname}{APPENDIX}
\section*{Appendices}
In these Appendixes, we provide detailed derivations about our results in the paper.
Appendix \ref{AppA} is devoted to supply algebraic-geometry tools which are invariant under stochastic local operation and classical communication (SLOCC). We write them for generic multipartite systems, unless otherwise specified. In Appendix \ref{AppB}, we provide a theorem about two-multilinear ranks for four-qubit systems and a Hasse diagram which helps in understanding the figure of petal-like classification of SLOCC orbits of four-qubit states in the paper. Finally, in Appendix \ref{AppC}, to show the effectiveness of our classification method, we provide an entanglement classification of five-qubit systems in terms of the families and subfamilies where one can easily discover the classifications of two-, three-, and four-qubit entanglement as the core structures, and hence, the interpretation of Mendeleev table.

\section{ALGEBRAIC-GEOMETRY TOOLS AND SLOCC INVARIANTS}\label{AppA}
Although it is customary to look at an $n$-partite quantum state
\begin{equation}\label{n-partite}
|\psi\rangle=\sum_{\alpha=1}^{n}\sum_{i_{\alpha}=0}^{d_{\alpha}-1}\mathbf{c}_{i_{1}\cdots i_{n}}|i_{1}\rangle\otimes\cdots\otimes|i_{n}\rangle
\end{equation}
as a vector, such a vector results from the vectorization of an order-$n$ tensor in the Hilbert space $\mathcal{H}_{n}=\otimes_{i=1}^{n}\mathbbm{C}^{d_{i}}$. In multilinear algebra, this vectorization is a kind of tensor reshaping. Here, we shall use a tensor reshaping known as tensor flattening (or matricization) \cite{Landsberg}. It consists in partitioning the $n$-fold tensor product space (here, $\mathcal{H}_{n}$) to two-fold tensor product spaces with higher dimensions. With respect to the partitioning, we define an ordered $\ell$-tuple $I=(i_{1},i_{2},\ldots,i_{\ell})$ where $1\leq\ell\leq{n-1}$ and $1\leq i_{1}<\cdots<i_{\ell}\leq{n}$ and an ordered $(n-\ell)$-tuple related to complementary partition $\bar{I}$ such that $I\cup\bar{I}=(1,2,\ldots,n)$. Therefore, $\mathcal{H}_{n}\simeq\mathcal{H}_{I}\otimes\mathcal{H}_{\bar{I}}$ where $\mathcal{H}_{I}=\otimes_{\alpha=i_{1}}^{i_{\ell}}\mathbbm{C}^{d_{\alpha}}$ and $\mathcal{H}_{\bar{I}}$ is the complementary Hilbert space.
Using Dirac notation, the matricization of $|\psi\rangle$ reads $\mathcal{M}_{I}[\psi]=\left(\langle{e_1}|\psi\rangle,\ldots,\langle{e_{d_{I}}}|\psi\rangle\right)^{\rm{T}}$, where $\{|e_j\rangle=|i_{1}\cdots{i_{\ell}}\rangle\}_{j=1}^{d_I=\Pi{d_{\alpha}}}$ is the computational basis of $\mathcal{H}_{I}$ and $\rm{T}$ denotes the matrix transposition.
Clearly, we shall consider all ordered $\ell$-tuples $I$ to avoid overlapping of entanglement families \cite{GM18}. Hence, for a given $|\psi\rangle$ we have as many matrix representations $\mathcal{M}_{I}[\psi]$ as the number of possible $\ell$-tuples $I$, which is ${\binom{n}{\ell}}$. In this way, we can define $\ell$-multilinear rank (hereafter $\ell$-multirank) \cite{Landsberg} of $|\psi\rangle$ as a ${\binom{n}{\ell}}$-tuple of ranks of $\mathcal{M}_{I}[\psi]$. Obviously, the zero-multirank is just a number, namely 1, as well as the $n$-multirank. Interestingly, we can see that the rank of $\mathcal{M}_{I}[\psi]$ is the same as the rank of the reduced density matrix obtained after tracing over the parties identified by the $(n-\ell)$-tuple $\bar{I}$, i.e., $\varrho_{I}={\rm{Tr}}_{\bar{I}}\left(|\psi\rangle\langle\psi|\right)=\mathcal{M}_{I}[\psi]\mathcal{M}_{I}^{\dagger}[\psi]$. The most important thing is that SLOCC equivalent states, i.e., $|\tilde{\psi}\rangle = \left(\otimes_{i=1}^{n} A_{i}\right) |\psi\rangle$, where $|\psi\rangle\in\mathcal{H}_{n}$ and $A_{i}\in{\rm{SL}}(d_{i},\mathbbm{C})$, yield $\mathcal{M}_{I}[\tilde{\psi}]=\left(\otimes_{i\in I}A_{i}\right)\mathcal{M}_{I}[\psi]\left(\otimes_{i\in\bar{I}}A_{i}\right)^{\rm{T}}$. Therefore, $\ell$-multirank is an invariant under SLOCC.

\begin{manualremark}{2}\label{remark2}
A state is genuinely entangled iff all $\ell$-multiranks are greater than one.
\end{manualremark}

For the case that each party has the same dimension, it is enough to check $\ell$-multiranks for partition $I$ with \mbox{$1\leq\ell\leq\lfloor\frac{n}{2}\rfloor$}, because for complementary partition $\bar{I}$ the \mbox{matrices} $\mathcal{M}_{\bar{I}}[\psi]$ are just the transpose of $\mathcal{M}_{I}[\psi]$ and transposition does not alter the rank of the matrix. For the multiqubit case, the order of such matrices can be from $2 \times 2^{n-1}$ to $2^{\lfloor\frac{n}{2}\rfloor} \times 2^{\lceil\frac{n}{2}\rceil}$ and the number of these matrices is the same as the number of possible $\ell$-tuples $I$ which ranges from ${\binom{n}{1}}$ to $(1/2)^{n+1~{\rm{mod}}~2} {\binom{n}{\lfloor\frac{n}{2}\rfloor}}$.

Since $\ell$-multiranks only depend on the state vector and, furthermore, because statements about rank can be rephrased as statements about minors which are determinants, it follows that a given $\ell$-multirank configuration determines a determinantal variety in the projective Hilbert space and multipartite pure states which have $\ell$-multiranks bounded by a given integer sequence make a subvariety of $\mathbbm{P}(\mathcal{H}_{n})$.
Indeed, these determinantal varieties are subvarieties of secant varieties of the projective variety of fully separable states. For a multipartite quantum state, the space of fully separable states can be defined as the Segre variety \cite{Miyake03, Heydari08}. The Segre embedding is
\begin{equation}\label{segre-SM}
\Sigma^{n}_{\textbf{d-1}}:~\mathbbm{P}^{d_{1}-1}\times\mathbbm{P}^{d_{2}-1}\times\cdots\times\mathbbm{P}^{d_{n}-1}\hookrightarrow\mathbbm{P}^{D}\,,
\end{equation}
where $\textbf{d-1}=(d_{1}-1,\ldots,d_{n}-1)$, $D=\left(\Pi_{i=1}^{n}d_{i}\right)-1$, and $\times$ is the Cartesian product of sets. One can easily check that $\Sigma$ is the projective variety of fully separable states. Indeed, if all partial traces give pure states, the corresponding ranks are all one. Conversely, if all $\ell$-multiranks are one, the state is fully separable. It is worth noting that multipartite symmetric separable states with identical parties of dimension $d$ have the structure of Veronese variety. The Veronese embedding is
\begin{equation}
\mathcal{V}^{n}_{d-1}:~\mathbbm{P}^{d-1}\hookrightarrow\mathbbm{P}^{m}\,,
\end{equation}
where $m={\binom{n+d-1}{d-1}}-1$.

Let projective varieties $\mathcal{X}$ and $\mathcal{Y}$ be subvarieties of a projective variety. The joining of $\mathcal{X}$ and $\mathcal{Y}$ is given by the algebraic closure, for the Zariski topology, of the lines from one to the other,
\begin{equation}\label{joinvariety}
\mathfrak{J}(\mathcal{X},\mathcal{Y})=\overline{\bigcup_{x\in\mathcal{X},y\in\mathcal{Y},x\neq{y}}\mathbbm{P}^1_{xy}}\,,
\end{equation}
where $\mathbbm{P}^1_{xy}$ is the projective line that includes both $x$ and $y$. Suppose now $\mathcal{Y}\subset\mathcal{X}$ and let tangent star $\mathcal{T}^{\star}_{\mathcal{X},\mathcal{Y},y_{0}}$ denotes the union of $\mathbbm{P}_{\star}^{1}=\lim_{x,y\to{y_{0}}}\mathbbm{P}^{1}_{xy}$ with $y_{0}\in\mathcal{Y}$. The variety of relative tangent star is defined as follows
\begin{equation}\label{tangentvariety}
\mathcal{T}(\mathcal{X},\mathcal{Y})=\bigcup_{y\in\mathcal{Y}}\mathcal{T}^{\star}_{\mathcal{X},\mathcal{Y},y}\,.
\end{equation}
If $\mathcal{X}=\mathcal{Y}$, the joining is called the secant variety of $\mathcal{X}$, i.e., $\sigma(\mathcal{X})=\mathfrak{J}(\mathcal{X},\mathcal{X})$, and we denote the tangential variety as $\tau(\mathcal{X})=\mathcal{T}(\mathcal{X},\mathcal{X})$. In addition, the iterated join of $k$ copies of $\mathcal{X}$ is called the $k$-secant variety of $\mathcal{X}$. Hence, the secant varieties that we have mentioned above are given by the algebraic closure of the joining of the Segre variety and the immediately previous secant variety:
\begin{equation}\label{recursivesecant}
\sigma_{k}(\Sigma)=\mathfrak{J}\left(\sigma_{k-1}(\Sigma),\Sigma\right)\,.
\end{equation}
Notice that the first secant variety of Segre variety coincides with the Segre variety itself, i.e., $\sigma_{1}(\Sigma)=\Sigma$. This means that a generic point of the $k$-secant is the superposition of $k$ fully separable states, whence we say that the generic tensor rank is $k$. We can also generalize the definition of tangent line to a curve by introducing its osculating planes \cite{Harris}. Hence, one can define varieties of different types of limiting curves inside the $k$-secant variety. To simplify the calculations, let $x_t$ be a smooth curve in $\Sigma$. Then, to get higher order information, we can take higher order derivatives and calculate the higher dimensional tangential varieties as follows:
\begin{equation}\label{G-tangent}
\tau_{k}(\Sigma)=\overline{\{x_{0}+x'_{0}+\cdots+x^{(k-1)}_{0}|x_{t}\subset\Sigma~\text{is a smooth curve}\}}\,.
\end{equation}
Obviously $\tau_{k}(\Sigma)\subset\sigma_{k}(\Sigma)$ and $\mathcal{T}(\tau_{k-1}(\Sigma),\Sigma)\subset\tau_{k}(\Sigma)$, the last inclusion is even an equality.

To obtain the dimension of the secants and tangents, one can utilize the following theorem \cite{Zak}.

\begin{manualtheorem}{1}\label{theorem1}
Let $\mathcal{X}\subset\mathbbm{P}^{D}$ be an irreducible nondegenerate (i.e., not contained in a hyperplane) $n$-dimensional projective variety. For an arbitrary nonempty irreducible $m$-dimensional variety $\mathcal{Y}\subset\mathcal{X}$ it is either ${\rm{dim}}~\mathfrak{J}(\mathcal{X},\mathcal{Y})=m+n+1 > {\rm{dim}}~\mathcal{T}(\mathcal{X},\mathcal{Y})=m+n$, or $\mathfrak{J}(\mathcal{X},\mathcal{Y})=\mathcal{T}(\mathcal{X},\mathcal{Y})$.
\end{manualtheorem}

Moreover, since the algebraic closure of the $\ell$-multirank is known to be the subspace variety \cite{Landsberg}, as mentioned in the paper, we have the following corollary.

\begin{manualcorollary}{1}\label{corollary}
$\ell$-multiranks of a given tensor in the $k$-secant are at most $k$.
\end{manualcorollary}

If the points of variety $\mathcal{X}$ remains invariant under the action of a group $G$, then so is any of its auxiliary variety which is built from points of $\mathcal{X}$. It means that the $k$-secant variety of Segre variety is invariant under the action of projective linear group and therefore is a SLOCC invariant. That is why the Schmidt rank, which indeed is tensor rank, is a SLOCC invariant. On the other hand, since tangent lines can be seen as the limits of the secant lines, there exist asymptotic SLOCC equivalence between two different SLOCC classes and, hence, we can find exceptional states as defined in Ref. \cite{ST13}.

To distinguish the elements of higher secants with the same $\ell$-multiranks, one can think about $\mathfrak{m}$ copies of projective Hilbert space and utilize $\mathfrak{m}^{\rm{th}}$ Veronese embedding, i.e.,
\begin{equation}\label{m-veronese}
\mathcal{V}_{D}^{\mathfrak{m}}:\mathbbm{P}(\mathcal{H}_{n})\to\mathbbm{P}({\rm{Sym}}^{\mathfrak{m}}[\mathcal{H}_{n}])\,,
\end{equation}
where ${\rm{Sym}}^{\mathfrak{m}}[\mathcal{H}_{n}]$ is the $\mathfrak{m}^{\rm{th}}$ symmetric power of Hilbert space $\mathcal{H}_{n}$ (${\rm{Sym}}^{\mathfrak{m}}[\mathcal{H}_{n}]\sim{\rm{Sym}}[\mathcal{H}_{n}^{\otimes\mathfrak{m}}]$). According to this embedding, one can use minors of catalecticant matrices \cite{LO13}, to find the elements of higher secants. Although, in principle, the minors of catalecticant matrices from Eq. (\ref{m-veronese}) provide us the invariant homogeneous polynomials, one can devise a more effective method. One of these, similar to the spirit of Ref. \cite{OS16}, could be based on projective invariants via an interpolation of representation theory \cite{Ottaviani13}. As we know, minors of catalecticant matrices are determinantal varieties and are invariant under the action of group $G={\rm{SL}}(d_{1},\mathbbm{C})\times\cdots\times{\rm{SL}}(d_{n},\mathbbm{C})$. Here, we should similarly provide homogeneous polynomials of degree $\mathfrak{m}$ which are invariant under the action of group $G$.
Given complex vector spaces $V_1\equiv\mathbbm{C}^{d_1},\ldots, V_n\equiv\mathbbm{C}^{d_n}$, the group $G$ acts over the tensor space ${\mathcal{H}}_{n}=\otimes_{i=1}^{n}V_i$ and, hence, on the polynomial ring,
\begin{equation}\label{polyring}
S=\sum_{\mathfrak{m}\ge 0}\mathrm{Sym}^{\mathfrak{m}}\left[{\mathcal{H}}_{n}\right]\,,
\end{equation}
where ${\mathcal{H}}_{n}^{\otimes\mathfrak{m}}\cong\left(V_{1}^{\otimes\mathfrak{m}}\right)\otimes\cdots\otimes\left(V_{n}^{\otimes\mathfrak{m}}\right)$. Since $G$ is a reductive group, every summand of degree $\mathfrak{m}$ of $S$ in Eq. (\ref{polyring}) decomposes as the sum of irreducible representations of $G$, which have the form $\otimes_{i=1}^{n}\mathfrak{S}_{\lambda_i}V_i$ for certain Young diagrams $\lambda_1,\ldots, \lambda_{n}$, each representation occurring with a multiplicity $m_{\lambda_{1}\cdots\lambda_{n}}$. When each $\lambda_i$ has a rectangular shape, with exactly $\dim{V_i}=d_{i}$ rows, all of the same length, we get that $\dim{\otimes_{i=1}^{n}\mathfrak{S}_{\lambda_i}V_i}=1$ and a generator of this space is known to be an invariant of degree $\mathfrak{m}$ and, indeed, all invariants occur in this way. In addition, these one-dimensional subspaces fill altogether the invariant subring $S^G$ of $S$, consisting of all invariant polynomials. It is known that such an invariant ring is finitely generated and in principle its generators and relations can be computed \cite{GoodmanWallach}. Note that the ideal of any $G$-invariant subvariety of the projective space $\mathbbm{P}(\mathcal{H}_{n})$, like the secant varieties, is generated by the generators of a finite number of summands of the form $\otimes_{i=1}^{n}\mathfrak{S}_{\lambda_i}V_i$. These subspaces are generally known as covariants, so an invariant is a covariant of dimension one, generated by a single $G$-invariant polynomial. A special case is given by codimension one $G$-invariant subvarieties of the projective space $\mathbbm{P}(\mathcal{H}_{n})$. Their ideal is principal and it is generated by a single invariant polynomial. Since the equations of any $k$-secant variety can be found among the $G$-covariants, which are invariant sets of polynomials, we give an explicit definition of a covariant and basic tools for constructing a complete set of covariants.

The $n$-partite state $|\psi\rangle$ in Eq. (\ref{n-partite}) can be interpreted as an $n$-linear form:
\begin{equation}\label{n-linear}
f({\bf{x}}^{1},\ldots,{\bf{x}}^{n})=\sum_{\alpha=1}^{n}\sum_{i_{\alpha}=0}^{d_{\alpha}-1}\mathbf{c}_{i_{1}\cdots i_{n}}{x}^{1}_{i_{1}}\cdots{x}^{n}_{i_{n}}\,.
\end{equation}
A covariant of $f$ is a multi-homogeneous $G$-invariant polynomial in the coefficients $\mathbf{c}_{i_{1}\cdots i_{n}}$ and the variables ${\bf{x}}^{\alpha}=\{{x}^{\alpha}_{i_{\alpha}}\}_{\alpha=1}^{n}$.
To construct covariants, we move on from Gour and Wallach \cite{GW13} who write all possible $\rm{SL}$ invariant polynomials for the action of $G$ over ${\mathcal{H}_n}$, following Schur-Weyl duality. Let $P_{d,\mathfrak{m}}$ denote the orthogonal projection of $\otimes^{\mathfrak{m}}\mathbbm{C}^d$ onto 
 $(\otimes^{\mathfrak{m}}\mathbbm{C}^d)^{{\rm{SL}}(d,\mathbbm{C})}$. Then, $P(v)=(P_{d_1,\mathfrak{m}}\otimes\cdots\otimes P_{d_n,\mathfrak{m}}(v^T))^T$, where $T$ stands for the intertwining map defined in Ref. \cite{GW13}, is the orthogonal projection from $\otimes^{\mathfrak{m}}{\mathcal{H}_n}$ to $(\otimes^{\mathfrak{m}}{\mathcal{H}_n})^G$. To compute $P_{d,\mathfrak{m}}$, first observe that it is zero if $\mathfrak{m}/d\notin{\mathbbm{Z}}$, while if $\mathfrak{m}=dr$ denote by $\chi_{d,r}$ the character of ${\mathfrak{S}}_\mathfrak{m}$ corresponding to the partition $\mathfrak{m}=r+\cdots+r$, and we get up to scalar multiples
\begin{equation}\label{inv-p}
P_{d,\mathfrak{m}}=\frac{d_{d,r}}{\mathfrak{m}!}\sum_{\pi\in{\mathfrak{S}}_{\mathfrak{m}}}\chi_{d,r}(\pi)\pi\,,
\end{equation}
where $d_{d,r}$ is the dimension of the irreducible representation corresponding to the partition $\mathfrak{m}=r+\cdots+r$ that can be calculated by the hook-length formula. This construction can be generalized to write all covariants of the above action, an invariant being a covariant of dimension $1$ as mentioned before. Every covariant of degree $\mathfrak{m}$ corresponds to $\otimes_{i=1}^{n}\mathfrak{S}_{\lambda_i}V_i$ for certain partitions $\lambda_i$ of $\mathfrak{m}$. Denoted by $\chi_{\lambda_i}$ the character of $\mathfrak{S}_{\mathfrak{m}}$ corresponding to the partition $\lambda_i$, we get again that up to scalar multiples,
\begin{equation}\label{cov-p}
P_{\lambda_i}=\frac{d_{\lambda_i}}{\mathfrak{m}!}\sum_{\pi\in{\mathfrak{S}}_{\mathfrak{m}}}\chi_{\lambda_i}(\pi)\pi
\end{equation}
is the orthogonal projection from $\otimes^{\mathfrak{m}}V_{i}$ to the isotypical summand containing $\mathfrak{S}_{\lambda_i}V_{i}$, so the orthogonal projection from $\otimes^{\mathfrak{m}}{\mathcal{H}_n}$ to $\otimes_{i=1}^{n}\mathfrak{S}_{\lambda_i}V_i$ is $P(v)=(P_{\lambda_1}\otimes\cdots\otimes P_{\lambda_n}(v^T))^T$. The drawback of this construction is the difficulty to check in advance which $P_{\lambda_i}$ appear in a covariant of degree $\mathfrak{m}$, that is when $\otimes_{i=1}^{n}\mathfrak{S}_{\lambda_i}V_i$ comes from the subspace ${\mathrm{Sym}}^{\mathfrak{m}}[{\mathcal{H}_n}]\subset\otimes^{\mathfrak{m}}{\mathcal{H}_n}$,
this problem is known as plethysm. For example, the partition $4=2+1+1$ gives the projection in Eq. (\ref{cov-p}),
\begin{align}\nonumber
v_1\otimes v_2\,\otimes & \,v_3\otimes v_4 \mapsto
\\ \nonumber
&\frac{1}{8}\Big(3\, v_1\otimes v_2\otimes v_3\otimes v_4
\\ \nonumber
&~\quad-\sum_{\pi\in(12)}v_{\pi(1)}\otimes v_{\pi(2)}\otimes v_{\pi(3)}\otimes v_{\pi(4)}
\\ \nonumber
&~\quad+\sum_{\pi\in(1234)}v_{\pi(1)}\otimes v_{\pi(2)}\otimes v_{\pi(3)}\otimes v_{\pi(4)}
\\ \nonumber
&~\quad-\sum_{\pi\in(12)(34)}v_{\pi(1)}\otimes v_{\pi(2)}\otimes v_{\pi(3)}\otimes v_{\pi(4)}\Big)\,,
\end{align}
where $(12)$ is the conjugacy class containing the six simple swaps and so on for the other conjugacy classes.

For the ``symmetric'' systems, there is also another well-known process in mathematics literature to construct the complete set of covariants. To interpolate physics and mathematics literatures, for a symmetric multiqubit system, the set of covariants is actually the set of joint covariants of binary forms and similarly for a symmetric multiqudit system, the set of covariants is the set of joint covariants of $d$-ary forms. A general method for constructing a complete set of covariants is known as transvectants, which are based on Cayley's omega process and are basic tools for this aim \cite{Olver}. Here, we give the procedure of creating transvectants for symmetric multiqudit systems [$d_{\alpha}=d$ for all $\alpha$ in Eq. (\ref{n-linear})]. Let functions $f_{1},\ldots,f_{d}$ be forms in variable ${\bf{x}}=(x_1,\ldots,x_d)$, and tensor product notation $f_1\otimes\cdots\otimes f_d$ denotes the $d$-fold join product $f_{1}({\bf{y}}_{1})\cdots f_{d}({\bf{y}}_{d})$ (note that ${\bf{y}}_{\gamma}=(y_{\gamma,1},\ldots,y_{\gamma,d})$, $\gamma=1,\ldots,d$). The $d$-dimensional Cayley omega process is the $d^{\rm{th}}$-order partial differential operator:
\begin{equation}\label{cayley-omega}
\Omega_{\bf{x}}=\left|\begin{array}{ccc} \frac{\partial}{\partial{y_{1,1}}} & \cdots & \frac{\partial}{\partial{y_{d,1}}} \\ \vdots & \ddots & \vdots \\ \frac{\partial}{\partial{y_{1,d}}} & \cdots & \frac{\partial}{\partial{y_{d,d}}} \end{array}\right|\,.
\end{equation}
The $r^{\rm{th}}$ transvectant of functions $f_{1},\ldots,f_{d}$ is
\begin{equation}\label{transvectant}
\left(f_{1},\ldots,f_{d}\right)^{(r)}={\rm{tr}}~\Omega^{r}_{\bf{x}}(f_{1}\otimes\cdots\otimes f_{d})\,,
\end{equation}
where ${\rm{tr}}$ sets all variables equal, i.e., ${\bf{y}}_{1}=\cdots={\bf{y}}_{d}={\bf{x}}$. For instance, the first and second transvectants are known as the Jacobian determinant and polarized form of Hessian. Now, if functions $f_{1},\ldots,f_{d}$ are $n$-tuple forms in $n$ independent $d$-ary variables ${\bf{x}}^{1},\ldots,{\bf{x}}^{n}$, one can define a multiple transvectant for any $\vec{\jmath}=(j_{1},\ldots,j_{n})\in\mathbbm{N}^{n}$ as follows:
\begin{equation}\label{n-transvectant}
\left(f_{1},\ldots,f_{d}\right)^{(\vec{\jmath})}={\rm{tr}}\prod_{i=1}^{n}\Omega^{j_{i}}_{{\bf{x}}^{i}}(f_{1}\otimes\cdots\otimes f_{d})\,.
\end{equation}
By building iterative tansvectants in the multigraded setting and starting with the covariant of degree 1, i.e., Eq. (\ref{n-linear}), one can provide a complete system of covariants for multiqudit systems. For instance, in Ref. \cite{BLT03} the complete set of covariants has been found for four-qubit systems with this method.

\section{MUCH ADO ABOUT TWO-MULTIRANKS FOR FOUR-QUBIT SYSTEMS}\label{AppB}
Carlini and Kleppe have classified all possible one-multiranks for any number of qudits \cite{CK11}. The case of two-multiranks is more subtle. The partial result of two-multiranks of four-qubit states which is related to the Fig. \ref{fig:1} can be seen in Hasse diagram in Fig. \ref{fig:2}. A partial classification was given classically in Ref. \cite{Segre}, where the case $(442)$ and its permutations were forgotten. The full classification is achieved by the following

\begin{figure}[t]
\center{\includegraphics[width=8.8cm]{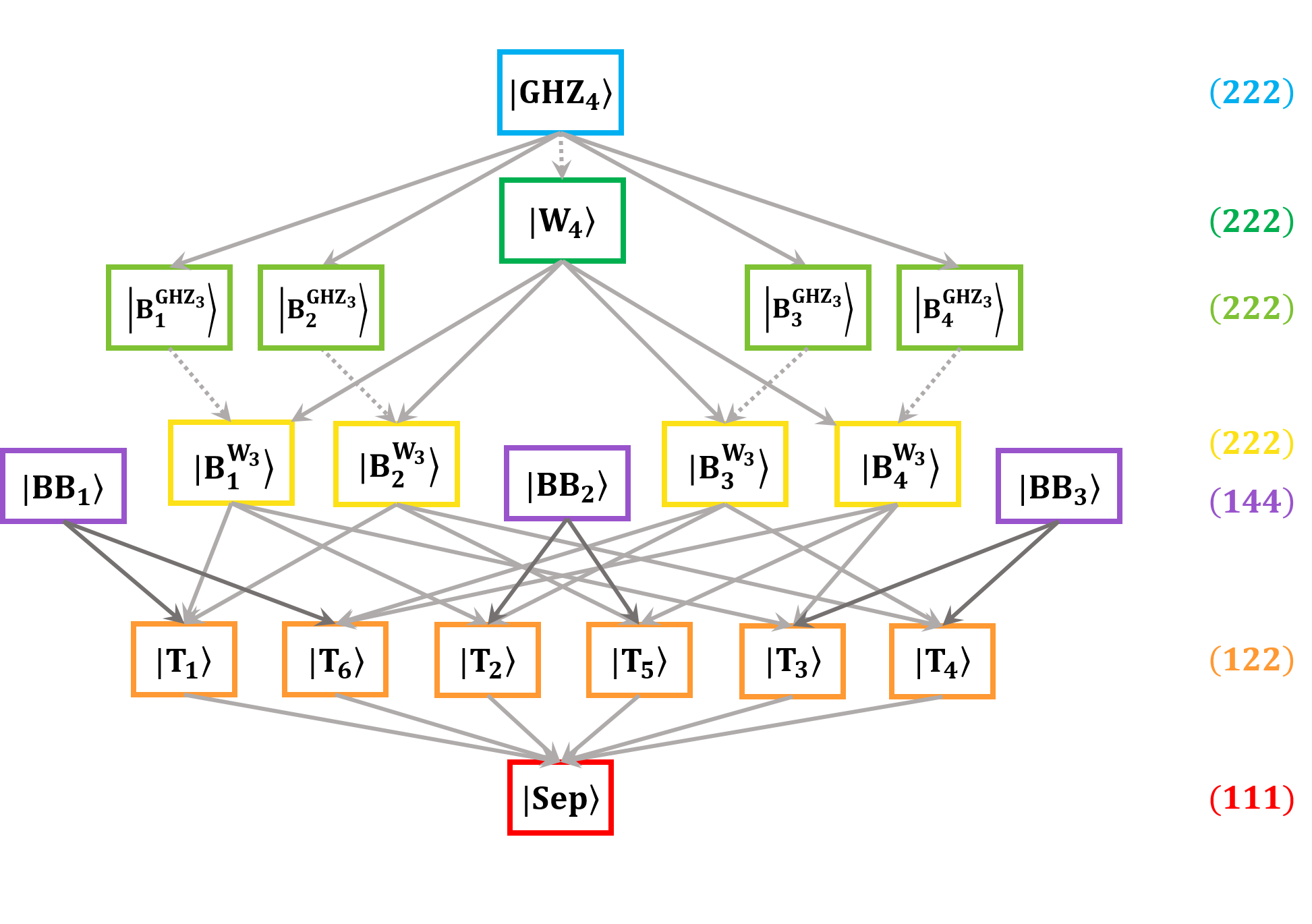}}
\caption{\label{fig:2} Hasse diagram of the central SLOCC classification of \mbox{four-qubit} states and their corresponding two-multiranks. The \mbox{arrows} denote noninvertible SLOCC transformations. When the arrow is dashed, the transformation is also approximated.}
\end{figure}

\begin{manualtheorem}{2}\label{theorem2}
(i) For any four-qubit system, the maximum among the three two-multiranks is attained at least twice. \\
(ii) The constraint in (i) is the only constraint for triples of two-multiranks of four-qubit systems, with the only exception of the triple $(133)$, which cannot be achieved.
\end{manualtheorem}

{\it Proof.}
If the minimum of the three two-multiranks is $\ge{3}$, the result follows from the fact that the three $4\times{4}$ determinants of the three flattenings sum to zero, as proved a century ago by Segre \cite{Segre}. Then, we assume that the minimum is $\le{2}$, attained by ${\mathcal{M}}_{xy}$ and we have three distinct cases as follows up to SLOCC [referring to Eq. (\ref{n-linear})]; here, multi-homogeneous coordinates for the four-qubit system are $x_iy_jz_kt_l$ for $i,j,k,l=\{0,1\}$).
\begin{enumerate}
\item[(1)] Secant:
$$f=x_0y_0(\sum a_{ij}z_it_j)+x_1y_1(\sum b_{ij}z_it_j)\,.$$
Here, the two-flattenings are $4\times 4$ matrices with the block form
$${\mathcal{M}}_{xz}=\left(\begin{array}{c|c}A&0 \\ \vspace{-3.5mm} \\ \hline \vspace{-3.5mm} \\ 0&B\end{array}\right)\,, \qquad {\mathcal{M}}_{xt}=\left(\begin{array}{c|c}A^{T}&0 \\ \vspace{-3.5mm} \\ \hline \vspace{-3.5mm} \\ 0&B^{T}\end{array}\right)\,,$$
which have the same rank. If this rank is one, then $A=0$ or $B=0$ and $f$ is a decomposable tensor.
\item[(2)] Tangent:
$$f=x_0y_0(\sum a_{ij}z_it_j)+(x_0y_1+x_1y_0)(\sum b_{ij}z_it_j)\,.$$
The two-flattenings have the block form
$${\mathcal{M}}_{xz}=\left(\begin{array}{c|c}A&B \\ \vspace{-3.5mm} \\ \hline \vspace{-3.5mm} \\ B&0\end{array}\right)\,, \qquad {\mathcal{M}}_{xt}=\left(\begin{array}{c|c}A^{T}&B^{T} \\ \vspace{-3.5mm} \\ \hline \vspace{-3.5mm} \\ B^{T}&0\end{array}\right)\,,$$
which again have the same rank. If this rank is one then $B=0$ and $f$ is a decomposable tensor.
\item[(3)] Isotropic:
$$f=x_0y_0(\sum a_{ij}z_it_j)+x_0y_1(\sum b_{ij}z_it_j)\,.$$
Here ${\mathcal{M}}_{xy}$ has rank $1$ iff $a$ and $b$ are proportional.
The two-flattenings have the block form
$${\mathcal{M}}_{xz}=\left(\begin{array}{c|c}A&B \\ \vspace{-3.5mm} \\ \hline \vspace{-3.5mm} \\ 0&0\end{array}\right)\,, \qquad {\mathcal{M}}_{xt}=\left(\begin{array}{c|c}A^{T}&B^{T} \\ \vspace{-3.5mm} \\ \hline \vspace{-3.5mm} \\ 0&0\end{array}\right)\,,$$
which have both rank $\le{2}$. If they have both rank one, then $A$ and $B$ are proportional, moreover $\mathrm{rk}(A)=\mathrm{rk}(B)=1$. This concludes the proof of (i). (ii) follows by exhibiting a representative for each case, as in Table \ref{table:2}. The nonexistence of case $(133)$ follows since when one two-multirank is $1$, then we may assume $f=(\sum a_{ij}x_iy_j)(\sum b_{ij}z_it_j)$
and depending on the pair $(\mathrm{rk}(A),\mathrm{rk}(B))=(1,1), (1,2), (2,2)$
we have, correspondingly, the triples $(111)$, $(122)$, $(144)$, so $(133)$ is not achieved.
\end{enumerate}
\qed

As for what concern the possibility of producing states in the lower secants and/or tangents from states in the higher secant and/or tangent by degeneration (Remark \ref{remark1}), from Fig. \ref{fig:2}, it results that we can asymptotically produce $|{\rm{W}}_4\rangle$ from $|{\rm{GHZ}}_4\rangle$ with a noninvertible SLOCC transformation, i.e., we cannot produce $|{\rm{GHZ}}_4\rangle$ from $|{\rm{W}}_4\rangle$. As a matter of fact, employing the singular (for $\epsilon\to{0}$) SLOCC transformation $A_{\epsilon}=\epsilon^{-1/4}\left(
\begin{array}{cc}
\sqrt[4]{-1} & 1 \\
\epsilon & 0 \\
\end{array}
\right)$, we get $\lim_{\epsilon\to0}A_{\epsilon}^{\otimes4}|\rm{GHZ}_4\rangle=|\rm{W}_4\rangle$.
Furthermore, based on Eq. (\ref{symmetric}), $|\rm{X}_{4}\rangle=\mathfrak{d}_{1}(|0001\rangle+|0010\rangle+|0100\rangle+|1000\rangle)+\mathfrak{d}_{4}|1111\rangle=\mathfrak{d}_{1}|\rm{W}_{4}\rangle+\mathfrak{d}_{4}|1111\rangle$ is a symmetric state in $\tau_{3}(\Sigma^{4}_{\textbf{1}})$ where $\mathfrak{d}_{0}=\mathfrak{d}_{2}=\mathfrak{d}_{3}=0$. It is obvious that if $\mathfrak{d}_4$ tends to zero we can approximately produce $|\rm{W}_4\rangle$ from $|\rm{X}_4\rangle$. As a matter of fact, employing the singular (for $\epsilon\to{0}$) SLOCC transformation $B_{\epsilon}=\epsilon^{-\frac{1}{4}}\left(
\begin{array}{cc}
 \sqrt[4]{-1} & (-1)^{7/12} 2^{2/3} \\
 \epsilon & 0 \\
\end{array}
\right)$, we can get $\lim_{\epsilon\to0}B_{\epsilon}^{\otimes4}|\rm{X}_4\rangle=|\rm{W}_4\rangle$.
As another example, employing the singular (for $\epsilon\to{0}$) SLOCC transformation $C_{\epsilon}=\epsilon^{-\frac{1}{4}}\left(
\begin{array}{cc}
 \sqrt[4]{-1} & \pm\sqrt{\frac{1}{2} \left(-\sqrt{3}-i\right)} \\
 \epsilon & 0 \\
\end{array}
\right)$, we can asymptotically produce $|\rm{W}_{4}\rangle$ from $|\rm{M}_4\rangle=\alpha|0000\rangle+\beta|0011\rangle+\gamma|1111\rangle$ belonging to $\sigma_{3}(\Sigma^{4}_{\textbf{1}})$, i.e., $\lim_{\epsilon\to0}C_{\epsilon}^{\otimes4}|\rm{M}_4\rangle=|\rm{W}_4\rangle$. It is also obvious that we can approximately produce $|\rm{GHZ}_4\rangle$ from $|\rm{M}_4\rangle$ by letting $\beta$ go to zero.

\section{FIVE-QUBIT ENTANGLEMENT CLASSIFICATION}\label{AppC}
For five-qubit states, due to Remark \ref{remark2}, Corollary \ref{corollary}, and classification of two-, three-, and four-qubit states, we have (1) all quadriseparable states $|{\rm{Q}}_{i}\rangle_{i=1}^{10}$ from Eq. (\ref{general2}) are elements of $\sigma_{2}(\Sigma^{5}_{\textbf{1}})$, (2) all triseparable states $|{\rm{T}}_{i}^{{\rm{GHZ}}_{3}}\rangle_{i=1}^{10}$ and $|{\rm{T}}_{i}^{{\rm{W}}_{3}}\rangle_{i=1}^{10}$ from Eq. (\ref{general3}) are, respectively, elements of $\sigma_{2}(\Sigma^{5}_{\textbf{1}})$ and $\tau_{2}(\Sigma^{5}_{\textbf{1}})$, and (3) all biseparable states $|{\rm{B}}_{i}^{{\rm{GHZ}}_{4}}\rangle_{i=1}^{5}$ and $|{\rm{B}}_{i}^{{\rm{W}}_{4}}\rangle_{i=1}^{5}$ from Eq. (\ref{general3}) are, respectively, elements of $\sigma_{2}(\Sigma^{5}_{\textbf{1}})$ and $\tau_{2}(\Sigma^{5}_{\textbf{1}})$. Considering Eq. (\ref{general3}), we can also find that states $|{\rm{GHZ}}_{5}\rangle$ and $|{\rm{W}}_{5}\rangle$ are elements of $\sigma_{2}(\Sigma^{5}_{\textbf{1}})$ and $\tau_{2}(\Sigma^{5}_{\textbf{1}})$, respectively. In a similar way to Eq. (\ref{general4}), all biseparable states of the form $|\sigma_{3}(\Sigma^{4}_{\textbf{1}})\rangle|\text{1-qubit}\rangle$ and $|\tau_{3}(\Sigma^{4}_{\textbf{1}})\rangle|\text{1-qubit}\rangle$ are elements of $\sigma_{3}(\Sigma^{5}_{\textbf{1}})$ and $\tau_{3}(\Sigma^{5}_{\textbf{1}})$, respectively. Note that the number of distinct subfamilies that these biseparable states create in each $\sigma_{3}(\Sigma^{5}_{\textbf{1}})$ and $\tau_{3}(\Sigma^{5}_{\textbf{1}})$, according to the permutations of the 1-qubit state, is, respectively, four times the number of subfamiles in $\sigma_{3}(\Sigma^{4}_{\textbf{1}})$ and $\tau_{3}(\Sigma^{4}_{\textbf{1}})$, i.e., $16$ subfamilies. Other elements of three-secant can be written in a similar way to Eq. (\ref{general4}) with a two-multirank including at least one 3 and no 4 (see Corollary \ref{corollary}). We denote these elements as $|(3\cdots)\rangle\in\sigma_{3}(\Sigma^{5}_{\textbf{1}})$ and $|(3\cdots)'\rangle\in\tau_{3}(\Sigma^{5}_{\textbf{1}})$. The remaining families of five-qubit states have different two-multiranks, including at least one 4. Considering classification of four-qubit as the core structure of five-qubit classification, all biseparable state of the form $|\sigma_{4}(\Sigma^{4}_{\textbf{1}})\rangle|\text{1-qubit}\rangle$ are elements of $\sigma_{4}(\Sigma^{5}_{\textbf{1}})$ ($40$ subfamilies). Here, we have a new type of biseparable state in five-qubit classification, i.e., $\mathcal{P}\{|{\rm{Bell}}\rangle|{\rm{GHZ}}_{3}\rangle\}$, which creates 10 subfamilies in $\sigma_{4}(\Sigma^{5}_{\textbf{1}})$ (see Table \ref{table:3}). Note that one can generate genuine entangled states from them which would be of the form $|{\rm{G}}_{5}^{2}\rangle$ ($\sim|{\rm{G}}_{5}^{3}\rangle$) in Eq. (\ref{general4}). On the limiting lines of these states, one can find the biseparable states $\mathcal{P}\{|{\rm{Bell}}\rangle|{\rm{W}}_{3}\rangle\}$ and the genuine entangled versions as the elements of $\tau_{4}(\Sigma^{5}_{\textbf{1}})$.
As another example, using reasoning similar to Eq. (\ref{tangent-dicke}), we can draw the following results for $n\geq 5$:
\begin{align}\label{general5}\nonumber
|{\rm{W}}_{n}\rangle+|1\rangle^{\otimes{n}}+\mathcal{P}\{|0\rangle^{\otimes{r}}|1\rangle^{\otimes{(n-r)}}\} &\in \tau_{4}(\Sigma^{n}_{\textbf{1}})\,, \\
|{\rm{D}}_{n}^{2}\rangle+\mathcal{P}\{|1\rangle^{\otimes{s}}|0\rangle^{\otimes{(n-s)}}\} &\in \tau_{4}(\Sigma^{n}_{\textbf{1}})\,,
\end{align}
where $2\leq{r}\leq{n-2}$ and $3\leq{s}\leq n-1$.

\begin{table*}[t]
\centering
\caption{Fine-structure classification of five-qubit entanglement (up to four-secant).}
\begin{tabular}{p{0.05\linewidth}p{0.1\linewidth}p{0.08\linewidth}p{0.19\linewidth}p{0.19\linewidth}p{0.19\linewidth}p{0.15\linewidth}}
\hline\hline
& & & & & & \\ [-2ex]
$\Sigma^{5}_{\textbf{1}}$ & $\sigma_{2}$ & $\tau_{2}$ & $\sigma_{3}$ & $\tau_{3}$ & $\sigma_{4}$ & $\tau_{4}$\\ [0.5ex]
\hline
& & & & & & \\ [-2ex]
$|{\rm{Sep}}\rangle$ & $|{\rm{GHZ}}_{5}\rangle$ & $|{\rm{W}}_{5}\rangle$ & $|(3333333333)\rangle$ & $|(3333333333)'\rangle$ & $|(4444444444)_{4}\rangle$ & $|(4444444444)'_{4}\rangle$ \\ [0.5ex]
& $|{\rm{B}}_{i}^{{\rm{GHZ}}_{4}}\rangle_{i=1}^{5}$ & $|{\rm{B}}_{i}^{{\rm{W}}_{4}}\rangle_{i=1}^{5}$ & $\vdots$ & $\vdots$ & $\vdots$ & $\vdots$ \\ [0.5ex]
& $|{\rm{T}}_{i}^{{\rm{GHZ}}_{3}}\rangle_{i=1}^{10}$ & $|{\rm{T}}_{i}^{{\rm{W}}_{3}}\rangle_{i=1}^{10}$ & $|(3\cdots)\rangle$ & $|(3\cdots)'\rangle$ & $|(4\cdots)_{4}\rangle$ & $|(4\cdots)'_{4}\rangle$ \\ [0.5ex]
& $|{\rm{Q}}_{i}\rangle_{i=1}^{10}$ & & $\mathcal{P}_{i}\{|\sigma_{3}(\Sigma^{4}_{\textbf{1}})\rangle|\text{1-qubit}\rangle\}_{i=1}^{16}$ & $\mathcal{P}_{i}\{|\tau_{3}(\Sigma^{4}_{\textbf{1}})\rangle|\text{1-qubit}\rangle\}_{i=1}^{16}$ & $\mathcal{P}_{i}\{|{\rm{Bell}}\rangle|{\rm{GHZ}}_{3}\rangle\}_{i=1}^{10}$ & $\mathcal{P}_{i}\{|{\rm{Bell}}\rangle|{\rm{W}}_{3}\rangle\}_{i=1}^{10}$ \\ [0.5ex]
& & & & & $\mathcal{P}_{i}\{|\sigma_{4}(\Sigma^{4}_{\textbf{1}})\rangle|\text{1-qubit}\rangle\}_{i=1}^{40}$ & \\ [0.5ex]
\hline\hline
\end{tabular}
\label{table:3}
\centering
\caption{Fine-structure classification of five-qubit entanglement (five- and six-secants).}
\begin{tabular}{p{0.17\linewidth}p{0.17\linewidth}p{0.17\linewidth}p{0.15\linewidth}}
\hline\hline
& & & \\ [-2ex]
$\sigma_{5}$ & $\tau_{5}$ & $\sigma_{6}$ & $\tau_{6}$ \\ [0.5ex]
\hline
& & & \\ [-2ex]
$|(4444444444)_{5}\rangle$ & $|(4444444444)'_{5}\rangle$ & $|(4444444444)_{6}\rangle$ & $|(4444444444)'_{6}\rangle$ \\ [0.5ex]
$\vdots$ & $\vdots$ & $\vdots$ & $\vdots$ \\ [0.5ex]
$|(4\cdots)_{5}\rangle$ & $|(4\cdots)'_{5}\rangle$ & $|(4\cdots)_{6}\rangle$ & $|(4\cdots)'_{6}\rangle$ \\ [0.5ex]
\hline\hline
\end{tabular}
\label{table:4}
\end{table*}

It is worth noting that since in the five-qubit case ($\mathbbm{C}^{2^{\otimes{5}}}$), we just have flattenings of sizes $2\times{16}$ and $4\times{8}$ with maximum ranks of 2 and 4, respectively, they do not provide nontrivial equations to find the elements of five-secant. Hence, with the method of Appendix \ref{AppA}, one can find, as in Ref. \cite{OS16}, homogeneous polynomials of degrees $6$ and $16$ where the rank of the Jacobian of these two equations gives the desired information (if the point is not singular for the five-secant then it cannot stay in the four-secant, i.e., it is an element of the proper five-secant family).

To have an exhaustive classification, we denote the other elements of four-, five-, and six-secants as $|(4\cdots)_{i\in\{4,5,6\}}\rangle\in\sigma_{i}(\Sigma^{5}_{\textbf{1}})$ and $|(4\cdots)'_{i\in\{4,5,6\}}\rangle\in\tau_{i}(\Sigma^{5}_{\textbf{1}})$ (see Tables \ref{table:3} and \ref{table:4}). It is worth noting that in the classification of five-qubit states, all the elements in five- and six-secant families are genuinely entangled.



\end{document}